\documentclass[smallextended]{svjour3}     
\smartqed  

\usepackage{graphicx}
\usepackage{epstopdf}
\usepackage{amsfonts}
\usepackage{float}
 \journalname{General Relativity and Gravitation}
\begin{document}

\title{Radiating fluid sphere immersed in an anisotropic atmosphere}

\author{N. F. Naidu$^\dag$, M. Govender$^\ddag$,
S. Thirukkanesh$^\mp$, S. D. Maharaj$^\ast$}%


\institute{N. F. Naidu, S. D. Maharaj \at Astrophysics and Cosmology Research Unit, School of Mathematics, Statistics and Computer Science, University of KwaZulu-Natal, Private Bag X54001, Durban 4000, South Africa. \\
\email{nolene.naidu@physics.org}      \email{maharaj@ukzn.ac.za}        \\
\and M. Govender \at Department of Mathematics, Faculty of Applied Sciences, Durban University of Technology, Durban, 4000, South Africa.\\
\email{megandhreng@dut.ac.za}\\
\and S. Thirukkanesh \at Department of Mathematics, Eastern University, Chenkalady, Sri Lanka. \\
\email{thirukkanesh@yahoo.co.uk} \\
\date{Received: date / Accepted: date}}
\maketitle
\begin{abstract}
We model a radiating star undergoing dissipative gravitational collapse in the form of radial heat flux. The exterior of the collapsing star is described by the generalised Vaidya solution representing a mixture of null radiation and strings. Our model generalises previously known results of constant string density atmosphere to include inhomogeneities in the exterior spacetime. By utilising a causal heat transport equation of the Maxwell-Cattaneo form we show that relaxational effects are enhanced in the presence of inhomogeneities due to the string density.
\end{abstract}

\section{Introduction}
Oppenheimer and Snyder \cite{snyder} pioneered research and interest in gravitational collapse with the examination of a dust sphere undergoing collapse. Spacetime singularities are regions in spacetime where densities and spacetime curvatures diverge. They are contained within general relativity, as a consequence of the mathematics involved in the theory. The Cosmic Censorship Conjecture \cite{pen}, states that any reasonable matter distribution that undergoes continued gravitational collapse will always form a black hole. However, there are several factors that can influence the outcome of such collapse. Exceptions have been proposed by researchers since interest in the study of late stage collapse has peaked \cite{josh,josh1,josh2,josh3}. \\

In 1951, Vaidya discovered a nonstatic solution of the Einstein field equations which describes the atmosphere of a radiating star \cite{Vaidya}. Later, Santos  derived the junction conditions to match any type of stellar interior to a radiating Vaidya exterior \cite{Santos}. To date, there have been many models of gravitational collapse under several sets of initial conditions and physical factors, derived from exact solutions to the Einstein field equations and appropriate versions of Santos' matching conditions. The Santos junction conditions were generalised to include the effects of an electromagnetic field and shearing anisotropic stresses during dissipative stellar collapse by de Oliveira and Santos \cite{olive}, and Maharaj and Govender \cite{sunil2}. Pinheiro and Chan  \cite{pin} examined shear-free nonadiabatic collapse in the presence of electric charge. The influence of pressure anisotropy, shear and bulk viscosity were also studied by Chan \cite{chan1,chan2}. Abebe et al. \cite{abebe1,abebe2} employed Lie symmetries to investigate the behaviour of radiating stars in conformally flat spacetime manifolds. Maharaj and Govender \cite{sunil3} studied gravitational collapse with isotropic pressure and vanishing Weyl stresses and showed that the stellar core was more unstable than the outer regions. Martinez \cite{martinez}, Herrera and Santos \cite{herrera}, Naidu et al. \cite{naidu}, and Naidu and Govender \cite{naidu2}, focussed on the thermodynamics of radiating stars and the importance of both the relaxation and mean collision time. \\

Maharaj et al. \cite{sunil1} showed the impact of the generalised Vaidya radiating metric on the junction conditions on the boundary of a radiating star. Their results describe a more general atmosphere surrounding the star, which is a superposition of the pressureless null dust and a string fluid. The string density was shown to affect the fluid pressure at the surface of the star. It was demonstrated that the string density reduces the pressure on the stellar boundary. The usual junction conditions for the Vaidya spacetime are regained in the absence of the string fluid. \\

Although several applications of irreversible thermodynamics in general relativity to date have employed Eckart theory, there are several shortcomings to this approach. In this theory, if a thermodynamic force is suddenly set equal to zero, then the corresponding thermodynamic flux vanishes instantaneously. This is a violation of relativistic causality, since it implies that the signal would propagate through the fluid at an infinite speed. This led to the development of causal theories of dissipative fluids (both relativistic and non-relativistic). There are several advantages to taking a causal approach: (i) There are causal propagations of dissipative signals for stable fluid configurations, (ii) There is no generic short-wavelength secular instability in causal theories, and (iii) The perturbations have a reasonably posed initial value problem, including the case of rotating fluids. Causal theories extend the space of variables of conventional theories by incorporating the dissipative quantities concerned (such as heat flux, particle currents, shear and bulk stresses). These physical quantities are treated similar to the conserved variables (such as energy density, particle numbers, etc.). This leads to a more comprehensive theory with a larger number of variables and parameters \cite{roy,kevin,robert}. \\

The layout of this paper is as follows: In $\S$2, we present the interior spacetime of a radiating star. In $\S$3, we focus on the exterior atmosphere. In $\S$4, we list the junction conditions for the smooth matching of the exterior and interior spacetimes. In $\S$5, we solve the junction condition and provide several classes of new exact solutions. Finally, our results are discussed in $\S$6. \\

\section{Conformally flat interior}
\label{sec:1}
The interior spacetime of the fluid sphere is represented by a
spherically symmetric, shear-free, conformally flat line element
\begin{eqnarray} \label{g1}
ds^2 &=& -\left(\frac{C_1 (t) r^2 +1}{C_2 (t) r^2 + C_3 (t)}\right)^2 dt^2 \nonumber\\
&& + \left(\frac{1}{C_2 (t) r^2 + C_3 (t)} \right)^2[ dr^2 +r^2 d\Omega^2],
\end{eqnarray}
where $d\Omega^2 = d\theta^2 +\sin^2 \theta d\phi^2$ and $C_1 (t)$, $C_2 (t)$ and $C_3 (t)$ are temporal functions yet to be determined \cite{conformally1}.  The matter content of the interior stellar region is described by an energy momentum tensor of a perfect fluid with heat flux
\begin{equation}\label{g2}
T_{ab} = (\mu + p)u_a u_b + p g_{ab} + q_a u_b
          + q_b u_a,
\end{equation}
where the energy density is represented by $\mu$, $p$ represents the pressure and the magnitude of the heat flux is $q = (q^aq_a)^{\frac{1}{2}}$. The comoving fluid four--velocity ${\bf u}$ has the form
\begin{equation}
u^a = \displaystyle\frac{C_2 (t) r^2 + C_3 (t)}{C_1 (t) r^2 +1} \delta^{a}_0. \label{2'}
\end{equation}
With the assumption that heat flow is in the radial direction and $ q^au_a = 0 $, the heat flow vector is given by
\begin{equation}
q^a = (0, q, 0, 0). \label{2''}
\end{equation}
The fluid collapse rate is given by $\Theta = u^a_{;a}$. We calculate this for our fluid sphere as
\begin{equation}
\Theta = -3 \frac{\dot{C_2}r^2 + \dot{C_3}}{C_1 (t) r^2 +1}. \label{2'''}
\end{equation}
The line element (\ref{g1}) has been extensively utilised by various authors to study dissipative collapse in which the Weyl stresses vanish within the stellar interior  \cite{conformally1,sunil3,conformally2,surisunil}. \\
In these investigations the line element is matched to Vaidya's outgoing solution which describes an isotropic and homogeneous atmosphere. The Weyl-free collapse model was subsequently utilised to model a radiating star with a two-fluid atmosphere \cite{megandhren}. The atmosphere consisted of a mixture of strings and null radiation. The density of the strings was assumed to be constant making the junction conditions mathematically tractable. When the string density vanishes the two-fluid atmosphere model is recovered \cite{sunil3}. Although the assumption of constant string density is highly simplified, it was shown that the presence of strings in the external spacetime leads to higher core temperatures. In this paper we relax the assumption of constant string density to incorporate inhomogeneity and anisotropy in the string density. \\
The boundary condition required for the smooth matching of the interior spacetime to the external generalised Vaidya solution is solved to produce an exact model of a radiating star with an inhomogeneous atmosphere. In the case of constant string density, our model reduces to the solution studied by Govender \cite{megandhren}. \\

For the line element (\ref{g1}) the Einstein field equations give
\begin{eqnarray} \label{field}
\mu &=& 3\left(\frac{{\dot C_2}r^2 + {\dot C_3}}{C_1r^2 + 1}\right)^2 + 12C_2C_3, \label{fielda}\\
p &=& \frac{1}{(C_1r^2 + 1)^2}\left[2({\ddot C_2}r^2 + {\ddot C_3})(C_2r^2 + C_3)\right. \nonumber\\
&&\left.-3({\dot C_2}r^2 + {\dot C_3})^2 - 2\frac{\dot C_1}{C_1r^2 + 1}({\dot C_2}r^2 + {\dot C_3})\right. \nonumber\\
&&\left.\times (C_2r^2 + C_3)r^2\right] + \frac{4}{C_1r^2 + 1} \nonumber\\
&&\times[C_2(C_2 - 2C_1C_3)r^2 + C_3(C_1C_3 - 2C_2)],\label{fieldb}\\
q &=& 4({\dot C_3}C_1 - {\dot C_2})\left(\frac{C_2r^2 +
C_3}{C_1r^2 + 1}\right)^2r.\label{fieldc}\end{eqnarray}
The temporal behaviour of the metric functions will be taken up in the next section.

\section{Exterior atmosphere}
The exterior spacetime is taken to be the generalised Vaidya solution which represents a two-fluid atmosphere made up of null radiation and a string fluid \cite{glass1}
\begin{eqnarray} \label{vaidya}
ds^2 &=& - \left(1 - 2\frac{ m(v,\sf{r})}{\sf{r}}\right)dv^2 -2 dvd{\sf{r}} \nonumber\\
&& + {\sf{r^2}} (d\theta^2 +\sin^2\theta d\phi^2),
\end{eqnarray}
where $m(v, \sf{r})$ is the mass function which represents the gravitational energy within a sphere of radius $\sf{r}$. The generalised Vaidya solution has been extensively studied within the context of gravitational collapse and Cosmic Censorship. Dawood and Ghosh \cite{dawood} found a class of nonstatic black hole solutions for a type-two fluid using the line element (\ref{vaidya}). They further showed that the end state of continued gravitational collapse is sensitive to the initial matter configuration. Recently, Mkenyeleye et al. \cite{naked1} studied the end states resulting from the gravitational collapse of a generalised Vaidya cloud. They showed that naked singularities are stable end states resulting from the collapse of a two-fluid composite such as null radiation and strings. \\

The energy momentum tensor consistent with the line element (\ref{vaidya}) is
\begin{equation}
T_{ab}^+ = \tilde{\mu} l_{a}l_{b}+\left(\rho+P\right)\left(l_{a}n_{b}+l_{b}n_{a}\right)+Pg_{ab},\label{gvem}
\end{equation}
where we have introduced two null vectors
\begin{eqnarray}
l_{a} &=& \delta_{a}^0, \label{gvemnv1}\\
\nonumber\\
n_{a} &=& \frac{1}{2}\left[1-2\frac{m(v,
\sf{r})}{\sf{r}}\right]\delta_{a}^{0}+\delta_{a}^{1},\label{gvemnv2}
\end{eqnarray}
and $l_{a}l^{a}=n_{a}n^{a}=0$ and $l_{a}n^{a}=-1$.
 The above energy momentum tensor describes a composite atmosphere made up of a superposition of pressureless null dust and a null string fluid \cite{husain,wang}.
In (\ref{gvem}) we interpret $\tilde{\mu}$ as the energy density of the null dust radiation, $\rho$ as the null string energy density, and $P$ as the null string pressure.
It is assumed that the strings diffuse, and that string diffusion is similar to point particle diffusion, i.e. that the number density diffuses from higher to lower numbers. Within this context, Glass and Krisch \cite{glass2} provided the following density profiles

\begin{eqnarray}\label{glass}
\rho &=& \rho_0 + k_1/\sf{r}, \label{glass1}\\
\nonumber\\
\rho &=& \rho_0 + k_3 v^{-3/2} \exp[-{\sf{r}}^2 /(4\mathcal{D} v)], \label{glass2}\\
\nonumber\\
\rho &=& \rho_0 + (k_4/{\sf{r}}) (1 + (\sqrt{\pi}/2) \mathrm{erf}[{\sf{r}}(4 \mathcal{D} v)^{-1/2}]), \label{glass3}
\end{eqnarray}
where $\rho_0$ is a constant. The first type represents free streaming radiation, the second represents diffusion and the third represents radiation. The radial pressure within the Vaidya atmosphere can be attributed to the string tension. In the standard Vaidya envelope with photons carrying energy away from the stellar core, Santos \cite{Santos} was able to show that the radial pressure is proportional to the heat flux at the boundary of the radiating star. This boundary condition determines the temporal evolution of the model. With the generalised Vaidya atmosphere it is the photons that carry energy to the exterior spacetime with the strings diffusing inwards. Our intention is to determine the effect of the string flux on physical parameters such as temperature and luminosity.

\section{Matching conditions}
\label{sec:2}
For a complete model of a radiating star the line element (\ref{g1}) is matched smoothly to the exterior spacetime (\ref{vaidya}) across a time-like hypersurface. The reader is referred to Maharaj et al. \cite{Sunil} for more insight into the junction conditions.

The mass profile of the collapsing sphere is given by
\begin{equation}
m(v, {\sf{r}})= \left[\frac{r^3\left[4(1 + C_1r^2)^2C_2C_3 + ({\dot C_2}r^2 + {\dot C_3})^2\right]}{2(1 + C_1r^2)(C_2r^2 + C_3)^3}\right]_{\Sigma},\label{stdmass}
\end{equation}
which is the total gravitational energy contained within the
stellar surface $\Sigma$. The continuity of the momentum flux
across the boundary of the star yields
\begin{equation}
p=\left[\frac{q}{C_2 (t) r^2 + C_3(t)}-\rho\right]_{\Sigma},\label{densres}
\end{equation}
which generalises the results obtained by Santos \cite{Santos}. It is clear from (\ref{densres}) that the pressure at the boundary of
the collapsing star depends on the magnitude of the heat flux $q$
and the exterior string density $\rho$. For our line element
(\ref{g1}) and the assumption of vanishing Weyl stresses,
(\ref{densres}) reduces to the nonlinear equation
\begin{eqnarray} \label{g9}
&& \ddot{C_2}b^2+\ddot{C_3}-\frac{3}{2}\frac{(\dot{C_2}b^2+\dot{C_3})^2}{C_2b^2+C_3}
-\frac{\dot{C_1}b^2(\dot{C_2}b^2+\dot{C_3})}{C_1b^2+1} \nonumber\\
&& -2(C_1\dot{C_3}-\dot{C_2})b +2\frac{(C_1b^2+1)}{C_2b^2+C_3}[C_2(C_2-2C_1C_3)b^2 \nonumber\\
&&+C_3(C_1C_3-2C_2)] \nonumber\\
&&+\frac{1}{2} \frac{(C_1b^2+1)^2}{C_2b^2+C_3}\left(\rho_0 + k_1\frac{C_2b^2+C_3}{b}\right)=0, \nonumber \\
\end{eqnarray}
where $r=b$ determines the boundary of the star. In the above we
have utilised the string density given in (\ref{glass1}). The string density chosen generalises previous attempts at modeling radiating stars with a generalised Vaidya atmosphere. Earlier works adopted a constant string density profile which made the junction condition (\ref{g9}) mathematically tractable. By setting $k_1 = 0$ we obtain previously known results for constant density string atmosphere \cite{Sunil}. We should also point out a subtle feature of the junction condition which is highlighted for the first time here. Recall that (\ref{densres}) is evaluated at the boundary ($r = r_\Sigma$). In addition, the continuity of the metric functions connects the interior radial coordinate to the exterior radial coordinate by
\begin{equation}
\left(\frac{r}{C_2 (t) r^2 + C_3 (t)}\right)_{\Sigma} = {\sf r}_\Sigma.
\end{equation}
Hence our choice of the string density profile $\rho = \rho_0 + k_1/{\sf r}$ implies that
\begin{equation} \label{whoa}
\rho =\rho_0 + \frac{k_1(C_2 (t) r^2 + C_3 (t)}{r},
\end{equation} at the {\it boundary}. This means that the string density profile (\ref{whoa}) generalises the constant string density profile to include anisotropy {\em and} inhomogeneity in the exterior since $\rho = \rho(r,t)$ at the boundary. This shows a direct link between the interior and exterior matter variables which is absent for the usual Vaidya metric with $m = m(v)$.

\section{Exact solution}
\label{sec:3}
In order to generate exact solutions of (\ref{g9}) we adopt the approach taken by Thirukkanesh et al. \cite{Moops}.
If we take
\begin{equation}
U(t) = C_1(t)b^2+1, \label{s1}
\end{equation}
then the governing equation can be written as
\begin{eqnarray}
&& (\dot{C_2}b^2+\dot{C_3}) \dot{U} +2 \left[\frac{\dot{C_3}}{b}-\frac{(C_2^2b^4-C_3^2)}{b^2(C_2b^2+C_3)} \right]U^2 \nonumber\\
&& \left[\frac{3}{2}\frac{(\dot{C_2}b^2+\dot{C_3})^2}{C_2b^2+C_3} -\frac{2}{b}(\dot{C_2}b^2+ \dot{C_3}) -(\ddot{C_2}b^2+\ddot{C_3})\right]U \nonumber\\
&& -\frac{1}{2(C_2b^2+C_3)} \left[4(C_3-2C_2b^2)\frac{C_3}{b^2}+{\rho}_0 \right. \nonumber\\
&& \left. +\frac{k_1}{b}(C_2b^2+C_3)\right]U^3 =0. \nonumber\\ \label{s2}
\end{eqnarray}
This equation has the generic structure
\begin{equation}
\mathcal{A}\dot{U}+\mathcal{B}U+\mathcal{C}U^2+\mathcal{D}U^3=0,
\label{s3}
\end{equation}
where we have set
\begin{eqnarray}
\mathcal{A} &=&\dot{C_2}b^2+\dot{C_3}, \nonumber\\
\mathcal{B}
&=&\frac{3}{2}\frac{(\dot{C_2}b^2+\dot{C_3})^2}{C_2b^2+C_3}
-\frac{2}{b}(\dot{C_2}b^2+\dot{C_3})-(\ddot{C_2}b^2+\ddot{C_3}),
\nonumber\\
\mathcal{C} &=&2 \left[
\frac{\dot{C_3}}{b}-\frac{(C_2^2b^4-C_3^2)}{b^2(C_2b^2+C_3)}\right],
\nonumber \\
\mathcal{D} &=& - \frac{1}{2(C_2b^2+C_3)}
\left[4(C_3-2C_2b^2)\frac{C_3}{b^2}+{\rho}_0 \right. \nonumber\\
&& \left. +\frac{k_1}{b}(C_2b^2+C_3)\right]. \nonumber
\end{eqnarray}
Equation (\ref{s3}) is an Abelian equation of the first kind in the
variable $U$.

We have the following class of solution
\begin{eqnarray}
C_1&=&\frac{1}{b^2}
\left(\frac{e^{2t/b}\dot{z}}{z^{3/2}
\left[K- \int \lambda \right]^{1/2}} -1  \right), \nonumber\\
&&\label{s7a}\\
C_2 &=& \frac{\dot{C_3}b \pm
\sqrt{\dot{C_3}^2b^2-4C_3(C_3+\dot{C_3}b)}}{2b^2}, \label{s7b}\\
C_3 &=& \mbox{arbitrary function}, \label{s7c}
\end{eqnarray}
where
\begin{eqnarray}
\lambda &=& \frac{e^{4t/b}\dot{z} \left(4(C_3-2C_2b^2)\frac{C_3}{b^2}+{\rho}_0 +\frac{k_1}{b}(z)\right)}{z^{4}}, \nonumber\\
z &=& C_2 b^2 + C_3,
\end{eqnarray}
and $K$ is a constant. It is interesting to note that this class of solution reduces to a particular category of the Misthry et al. \cite{surisunil} solution in the limit of vanishing string density (i.e. ${\rho}_0=k_1=0$).

\section{Physical analysis}
\label{sec:4}
Our aim is to determine the influence of an inhomogeneous and anisotropic atmosphere on the temperature distribution of the collapsing body. By utilizing a causal heat transport equation, it was shown that temperature is increased at each interior point of the collapsing star when the atmosphere is composed of a mixture of pure radiation and strings with uniform density \cite{megandhren}. The truncated causal transport equation for the line element (\ref{g1}) is given by \cite{megandhren}
\begin{eqnarray} \label{mm}
&& \beta\left(\frac{q}{C_2r^2 + C_3}\right)\,\!\raisebox{2mm}{$\cdot$} \,\,T^{-\sigma} + \frac{q(C_1r^2 + 1)}{(C_2r^2 + C_3)^2} \nonumber\\
&& = -\left(\frac{\alpha}{C_2r^2 + C_3}\right)\left(\frac{(C_1r^2 + 1)T}{C_2r^2 + C_3}\right)'T^{3 -\sigma}\,,
\end{eqnarray}
for the physically motivated choices of the thermal conductivity
$\kappa$, the mean collision time $\tau_c$, and the relaxation time $\tau$:
\begin{equation} \kappa =\gamma T^3{\tau}_{\rm c}, \hspace{1cm}
 \tau_{\rm c} =\left({\alpha\over\gamma}\right)
T^{-\sigma}, \hspace{1cm}\tau =\left({\beta\gamma \over
\alpha}\right) \tau_{\rm c} \,. \label{5c}\end{equation} The quantities
$\alpha \geq 0$, $\beta \geq 0$, $\gamma \geq 0$ and $\sigma \geq 0$ are constants.

For the special case of constant collision time which corresponds to $\sigma = 0$, it is possible to integrate (\ref{mm}) \cite{gov3}. The noncasual heat transport equation is recovered when we set $\beta = 0$ in (\ref{mm}). \\

\begin{figure*}
\includegraphics[width=0.85\textwidth]{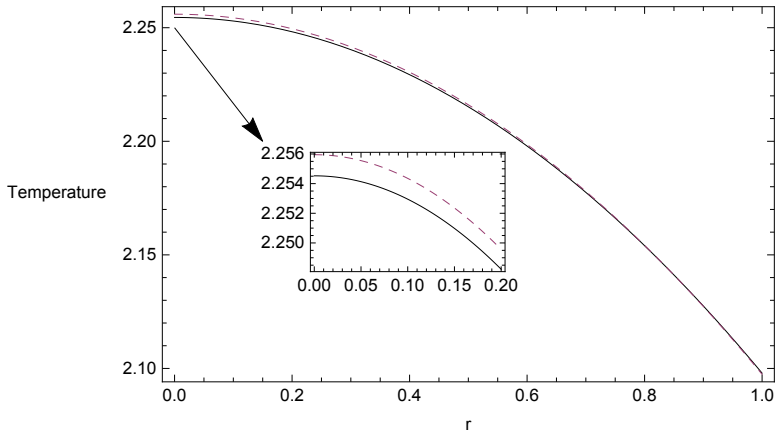}\caption{Non-causal temperature profile for constant density (solid line) and non-constant density (dashed line), with $\beta = 0$, $A_0 = 10$, $r_0 = 1$, $b = 10$, $K = 0.0003$, $\rho_0 = 1$, $t = -10$, and $k_1 = 0$ for non-constant density.} \label{fig1}
\end{figure*}

In order to investigate the behaviour of the temperature profiles in the presence of a cloud of anisotropic and inhomogeneous strings and null radiation we need to specify the arbitrary temporal functions. To this end we choose $C_3(t) = ae^{bt}$ in (\ref{s7a})--(\ref{s7c}) which ensures that the function $C_3(t)$ is continuous and free of singularities. With this choice, relation (\ref{s7b}) implies that $C_2(t) = C_3(t)$ and since $C_1(t)$ is defined in terms of $C_2(t)$, $C_3(t)$ and their derivatives, the temporal behaviour of our model is now fully specified. We are now in a position to obtain the causal temperature profile for the collapsing body. \\
Fig.~\ref{fig1} shows the Eckart temperature distribution as a function of the radial coordinate for both the isotropic string density ($\rho = {\mbox constant}$) and the nonuniform, inhomogeneous string atmosphere. Fig.~\ref{fig1} indicates that the temperature profiles are almost identical during the early stages of collapse. This is expected as the star is in quasi-static equilibrium and the anisotropy and inhomogeneity in the `young' atmosphere are indistinguishable from an isotropic and homogeneous atmosphere. As the collapse proceeds, anisotropic effects and density inhomogeneities become more prominent in the atmosphere. The result of this is a reduction of heat transfer to the exterior spacetime which results in a higher core temperature. We expect the central temperature to be significantly different to the surface temperature as the surface layers of the star are much cooler than the central region. \\

\begin{figure*}
\includegraphics[width=0.85\textwidth]{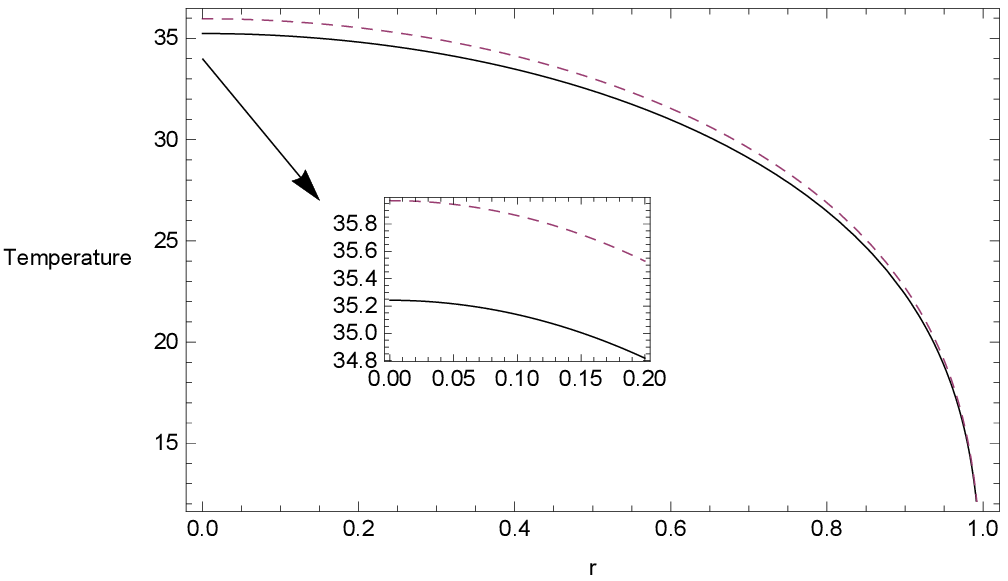}\caption{Causal temperature profile for constant density (solid line) and non-constant density (dashed line), with $\beta = 10^6$, $A_0 = 10$, $r_0 = 1$, $b = 10$, $K = 0.0003$, $\rho_0 = 1$, $t = -10^{-5}$, and $k_1 = 10000$ for non-constant density.} \label{fig2}
\end{figure*}

In Fig.~\ref{fig2} we have plotted the causal temperature profile ranging from the center of the star to the boundary. We observe that the causal temperature profile for the nonconstant density profile is higher at each interior point when compared to the uniform density profile. This difference is much higher at the core than at the surface layers. In addition, the difference in the causal temperature profiles is more accentuated than their noncausal counterparts closer to the center of the star. In observing Fig.~\ref{fig1} and Fig.~\ref{fig2}, we must bear in mind that that noncausal temperature is calculated for $\beta = 0$, when the star is close to equilibrium (early times). The causal temperature  is obtained for large values of $\beta$ which correspond to later times of the collapse process. The causality 'index' $\beta$ is intimately linked to the temporal evolution of the collapsing body. 

\begin{figure*}
\includegraphics[width=0.75\textwidth]{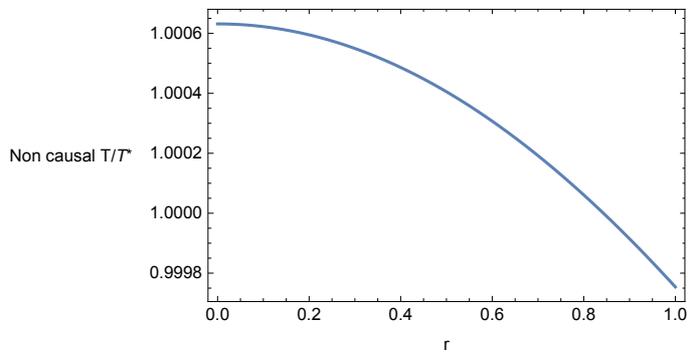}\caption{Ratio of Non-causal temperature profile: non-constant density over constant density with $t = -0.0001$.} \label{fig3}
\end{figure*}

\begin{figure*}
\includegraphics[width=0.75\textwidth]{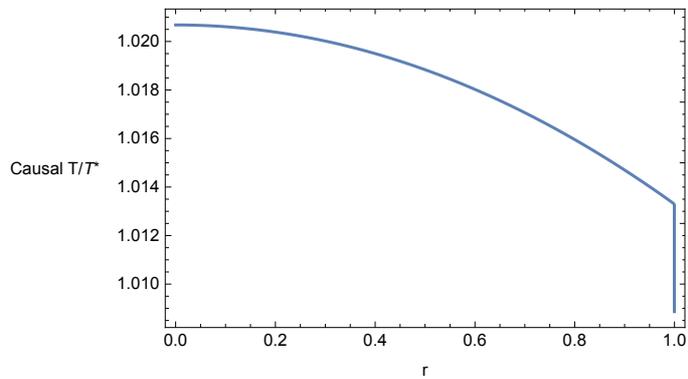}\caption{Ratio of Causal temperature profile: non-constant density over constant density with $t = -10$.} \label{fig4}
\end{figure*}

In order to determine the impact of the inhomogeneity we have plotted the ratio of the temperature $(T)$ arising from nonuniform density to the temperature $(T^{*})$ resulting from a uniform string density. The results are plotted in Fig.~\ref{fig3} (noncausal temperature) and Fig.~\ref{fig4} (causal temperature). We observe that inhomogeneity effects on the temperature profiles are much more marked at the core compared to the cooler surface layers. Fig.~\ref{fig3} shows that the ratio of $T/T^{*}$ is close to unity throughout the collapsing body. We expect this since during this epoch inhomogeneity effects in the exterior are small. As the collapse proceeds, there will be a steady flux of radiation from the core to the exterior thus enhancing inhomogeneity in the atmosphere. Fig.~\ref{fig4} shows the deviation of $T/T^{*}$ from unity. 

In this work we have modeled dissipative collapse of a radiating star in which the atmosphere is composed of null radiation and anisotropic string distribution. We were able to integrate the boundary condition for a nonuniform string density thus generalising previous investigations involving constant string densities. We showed that the nonuniform string density present in the atmosphere leads to higher core temperatures with the enhancement being more pronounced closer to the central regions of the collapsing body. A natural extension of this investigation is to include shearing effects within the collapsing core and to determine the subsequent evolution of the temperature profiles of the radiating star. \\

\begin{acknowledgements}
The authors are grateful to the National Research Foundation (NRF) and both the Durban Institute of Technology and University of KwaZulu-Natal. SDM acknowledges that this research is supported by the South African Research Chair Initiative of the Department of Science and Technology and the NRF.
\end{acknowledgements}

\label{lastpage}
\end{document}